\documentclass[aps,preprint,nofootinbib]{revtex4}
 \usepackage{graphicx}
 \def\be{\begin{equation}}
 \def\ee{\end{equation}}
 \newcommand{\lsim}{\,\raise-0.3ex\hbox{$\sim$}\kern-.7em\hbox{$^>$}\,}
\begin{document}

\title{From QCD lattice calculations to the equation of state of
    quark matter}

\author{A.~Peshier}
 \email{Andre.Peshier@theo.physik.uni-giessen.de}
 \affiliation{Institut f\"ur Theoretische Physik, Universit\"at Giessen,
    35392 Giessen, Germany}
\author{B.~K\"ampfer}
 \affiliation{Forschungszentrum Rossendorf, PF 510119, 01314 Dresden,
    Germany}
\author{G.~Soff}
 \affiliation{Institut f\"ur Theoretische Physik, TU Dresden,
    01062 Dresden, Germany}

\date{\today}

\begin{abstract}
 We describe two-flavor QCD lattice data for the pressure at nonzero
 temperature and vanishing chemical potential within a quasiparticle
 model. Relying only on thermodynamic consistency, the model is
 extended to nonzero chemical potential.
 The results agree with lattice calculations in the region of small
 chemical potential.
\end{abstract}


\maketitle

\section{Introduction}

One of the fundamental issues which triggered, and has influenced since,
heavy ion physics is the question of the phase structure and the
thermodynamic properties of strongly interacting matter at energy densities
above 1\,GeV/fm$^3$.
Under such conditions, exceeding the energy density in nuclei but still far
away from the asymptotic regime, the coupling strength $\alpha_s$ is large,
which makes the theoretical description of the many-body problem
challenging.

In the recent past the understanding of this field has become much
more detailed. The phase diagram for QCD with $n_{\!f} = 2$ massless
flavors, which is the case we will consider in the following, can be
briefly described as follows (we refer to \cite{RajaW} for a recent
review). At zero quark chemical potential, $\mu = 0$, the broken
chiral symmetry of hadron matter is restored within the quark-gluon
plasma, at a critical temperature $T_c \approx 170\,$MeV. It is
thought that this second order transition persists also for nonzero
$\mu$, thus defining a critical line, which changes to a first order
transition line at the tricritical point. For small temperatures and
$\mu \lsim \mu_c$ one anticipates a color-superconducting phase of
quark matter. The value of $\mu_c$ is expected to be 100...200\,MeV
larger than the quark chemical potential $\mu_n = 307\,$MeV in
nuclear matter. Quantitative results for large $\alpha_s$ can be
obtained from first principles by lattice calculations which were,
however, restricted to nonzero temperature and $\mu = 0$ until very
recently. Therefore, the described picture for $\mu \not= 0$ is
mainly based on general arguments combined with results from various
models, including extrapolations of perturbative QCD.

As a phenomenological description of the thermodynamics of deconfined
strongly interacting matter we proposed a quasiparticle model
\cite{PKPS96, PKS00}. Its parameters are fixed by the lattice data at
$\mu = 0$. We then use the fact that within the model the
thermodynamic potentials at zero chemical potential and $\mu \not= 0$
are related by thermodynamic consistency. In \cite{PKS00} we analyzed
lattice data for $n_{\!f}=2$ flavors \cite{Karsc}, and $n_{\!f}=4$
\cite{Engel}, which were, however, still derogated by sizable lattice
artifacts which have an effect on the absolute scaling of the data.
We therefore introduced a constant effective number of degrees of
freedom of the quasiparticles as an additional model parameter to
obtain first qualitative estimates. Later we considered in
\cite{PKS01} the lattice data \cite{Karsch00}, where also the
physical case of (2+1) flavors was simulated. As the absolute scaling
of the lattice data enters as important information in particular
near $T_c$, we pragmatically applied the continuum extrapolation of
the data, which was proposed in \cite{Karsch00} for $T > 2\, T_c$,
also for smaller temperatures. The results of this prescription can
now be compared to new lattice data \cite{CPPACS}. Meanwhile, there
are other lattice calculations which allow us to test directly the
assumptions underlying the quasiparticle model as well as, for the
first time, some of its predictions for nonzero chemical potential.

We will therefore consider here the presently available lattice data
for $n_{\!f}=2$. Based on that, we will fit and discuss the
quasiparticle parameters at $\mu=0$ in Section 3. In Section 4, we
will briefly summarize how to extend the model to nonzero chemical
potential, and compare our findings with the results
\cite{AlltHKKLSS} from lattice simulations studying the region of
small $\mu$.  Section 5 concludes with the discussion of some
physical implications.

\section{Finite temperature lattice data}

\begin{figure}[hbt]
  \centerline{\includegraphics[scale=1]{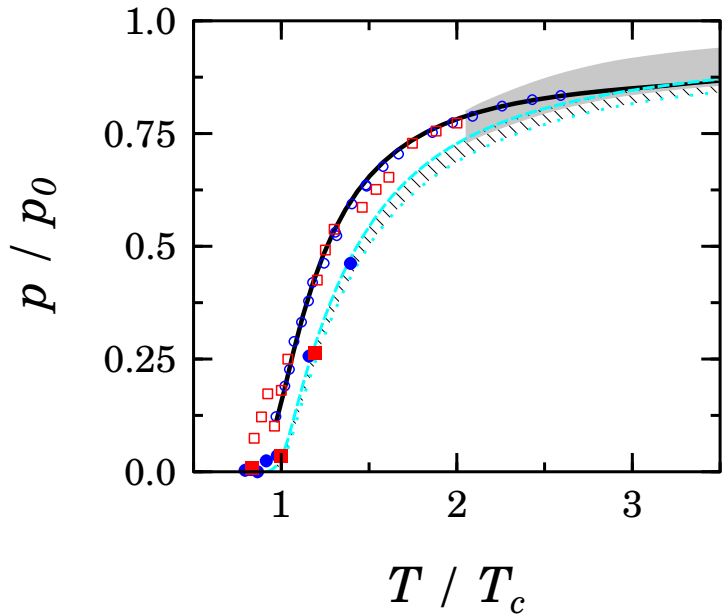}}
  \caption{Compilation of $n_{\!f}=2$ lattice data for the pressure in
    units of the free pressure $p_0$.
    Shown are the scaled (see text) data \cite{CPPACS} for light quarks
    corresponding to meson mass ratios of $0.65 \le m_{ps}/m_v \le 0.75$
    (open circles: $N_\tau=4$; open squares: $N_\tau=6$), and the
    continuum estimate \cite{Karsch00} (grey band).
    The full line is the quasiparticle result.
    The full symbols represent the data \cite{CPPACS} for large quark
    masses, with $m_{ps}/m_v = 0.95$.
    For comparison, the hatched band shows the SU(3) lattice data
    (dotted line: \cite{Boyd}; dashed line: \cite{Okamo}) normalized
    to the corresponding free pressure.
    \label{fig:latt dat p}}
\end{figure}

The simulations \cite{CPPACS} are performed on lattices with spatial
extent $N_\sigma=16$ and temporal sizes $N_\tau = 4$ and $N_\tau =
6$, with an improved Wilson quark action and renormalized quark
masses corresponding to fixed ratios $m_{ps}/m_v$ of the pseudoscalar
to vector meson masses.  We first consider the data for two light
flavors, corresponding to $0.6 \le m_{ps}/m_v \le 0.75$. Although
this is larger than the physical value, the results are almost
insensitive to the ratio, which suggests that they are not too far
from the chiral limit. As expected for the rather small lattice
sizes, the results for $N_\tau=4$ and 6 differ. However, we observe
that normalizing the pressure data by $p_0^{\rm cont}/p_0^{N_\tau}$,
the ratio of the free limits in the continuum and on the lattice,
improves considerably the consistency between the data sets. As a
matter of fact, the normalized $N_\tau=4$ data are in agreement with
the normalized $N_\tau=6$ data after rescaling by a constant of
$1.14$. This simple scaling behavior for large coupling is rather
remarkable. Based on this observation we suggest the continuum
estimate for the pressure shown in Fig.~\ref{fig:latt dat p}.
We assume here that the normalized $N_\tau=6$ data are already close
to the continuum limit. This is supported by the fact that the thus
interpreted data match the aforementioned continuum estimate from the
staggered quark simulations \cite{Karsch00}\footnote{
    In these calculations $m_q = 0.1\,T$ was assumed, corresponding to
    $m_{ps}/m_v = 0.7$ at $T_c$. From the weak quark mass sensitivity
    observed in \cite{CPPACS}, both results should indeed be comparable.}.
Therefore, a consistent picture forms for the thermodynamics of QCD with
$n_{\!f} = 2$ light flavors.

\begin{figure}[hbt]
  \centerline{\includegraphics[scale=1]{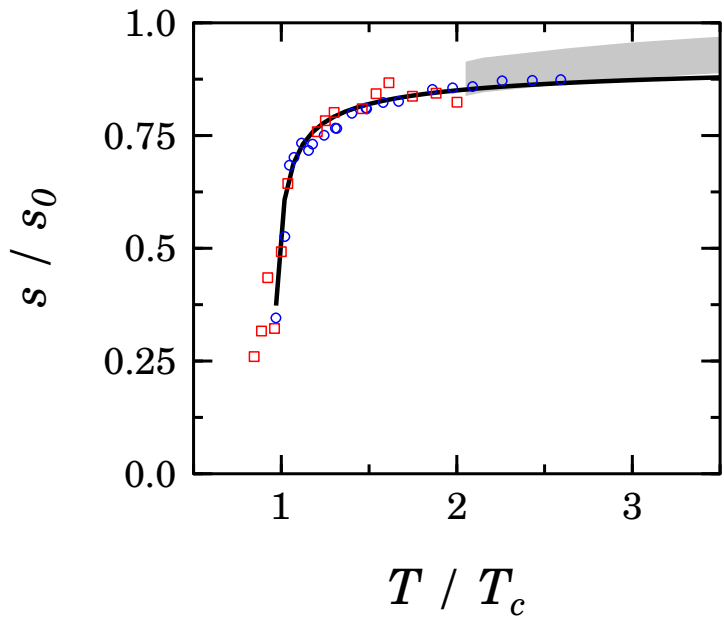}}
  \caption{The lattice data for the entropy corresponding to the data for
  the pressure shown in Fig.~\ref{fig:latt dat p}, and the quasiparticle
  fit.
  \label{fig:latt dat s}}
\end{figure}

In Fig.~\ref{fig:latt dat s}, the corresponding data for the entropy
are shown. It is noted that since the slope of the continuum
extrapolated pressure \cite{Karsch00} is slightly larger than that
from the data \cite{CPPACS} (see Fig.~\ref{fig:latt dat p}), the upper part
of the error band is already for $T \sim 3\, T_c$ very close to the free
limit.
This would be in contrast to the pure gauge case, where the uncertainty due
to lattice artifacts has become small, so we will assume that the lower side
of this estimate is more relevant.

\section{Quasiparticle model}

For completeness, we briefly recall here the main ideas of the
quasiparticle model \cite{PKPS96,PKS00} of the QCD plasma.

For weak coupling $g$, the thermodynamic behavior of the system is
dominated by its excitations with momenta $\sim T$. While hard
collective modes (the longitudinal plasmon and the quark hole
excitation) are exponentially suppressed, the transverse gluons and
the quark particle excitations propagate predominantly on simple mass
shells, $\omega_i^2(k) \approx m_i^2+k^2$ \cite{leBel}. In the chiral
limit the so-called asymptotic masses are given by
\begin{eqnarray}
  m_g^2
  &=&
  \frac16
  \left[
     \left( N_c+ \frac12\, n_{\!f} \right) T^2
   + \frac{N_c}{2\pi^2} \sum_q \mu_q^2
  \right] g^2 \, ,
  \nonumber \\
  m_q^2
  &=&
  \frac{N_c^2-1}{8N_c}\, \left[ T^2+\frac{\mu_q^2}{\pi^2} \right] g^2 \, ,
 \label{m2}
\end{eqnarray}
where $\mu_q$ denotes the quark chemical potential, and $N_c=3$.
Interpreting the relevant excitations as quasiparticles, the
thermodynamic potential is
\be
  p(T,\mu)
  =
  \sum_i p_i(T, \mu_i(\mu); m_i^2) - B(m_j^2) \, ,
  \label{p_eff}
\ee where $p_i = \pm d_i\,T \int d^3k/(2\pi)^3\, \ln(1 \pm
\exp\{-(\omega_i-\mu_i)/T\})$ are the contributions of the gluons
(with vanishing chemical potential) and the quarks (for the
antiquarks, the chemical potential differs in the sign), and
$d_g=2(N_c^2-1)$ and $d_q=2N_c$ count the degrees of freedom. As
shown in \cite{GoreY}, thermodynamic consistency requires the
derivative, with respect to the $m_j^2$, of the right-hand side of
Eq.~(\ref{p_eff}) to vanish, i.\,e., the contribution $B$ is related
to the $T$ and $\mu$ dependent masses by
\be
  \frac{\partial B}{\partial m_j^2}
  =
  \frac{\partial p_j(T,\mu_j; m_j^2)}{\partial m_j^2} \, .
 \label{B'}
\ee This implies that the entropy and the particle densities are
simply given by the sum of the individual quasiparticle
contributions,
\be
  s_i
  =
  \left.
   \frac{\partial p_i (T, \mu_i; m_i^2)}{\partial T}
  \right|_{m_i^2}
  \! , \quad
  n_i
  =
  \left.
   \frac{\partial p_i(T, \mu_i; m_i^2)}{\partial \mu_i}
  \right|_{m_i^2} \, ,
 \label{s,n}
\ee while the energy density has the form $e = \sum_i e_i + B$.

Expanded in the coupling $g$, the above approach reproduces the
leading-order perturbative results. The full expressions, however,
represent a thermodynamically consistent resummation of terms of all
orders in $g$. This suggests pondering the application of the model
also in the strong coupling regime.\footnote{
    A formal reason supporting this attempt is the stationarity of
    the thermodynamic potential with respect to variation of the
    self-energies around the physical value; see \cite{sca} and the
    references given there.
    Moreover, there are heuristic arguments that resummation improved
    leading order results might be more appropriate at large coupling
    than high order perturbative results \cite{Pesh98}.}
Considering first the case $\mu=0$, it indeed turns out that the
lattice data for the entropy shown in Fig.~\ref{fig:latt dat s} can
be described by the model with the ansatz
\be
 \alpha_s(T,\mu=0)
 =
 \frac{12\pi}{(11N_c-2n_{\!f})\ln[\lambda(T-T_s)/T_c]^2}
 \label{eq:alpha(T)}
\ee for $g^2/(4\pi)$. This is the leading order perturbative result
at a momentum scale determined by the temperature: $T_c/\lambda$ is
related to the QCD scale $\Lambda$, while $T_s$ parameterizes the
behavior in the infrared. For the parameters we obtain\footnote{
    For the fit we considered only the normalized data \cite{CPPACS}.
    The result then reproduces the extrapolated data \cite{Karsch00}
    on the lower side of the estimated error band; see the remark
    at the end of the last section.}
\be
  \lambda = 17.1 \, , \qquad
  T_s = 0.89\, T_c \, .
\ee The resulting quasiparticle masses are large; near $T_c$ they
reach several times the value of the temperature.\footnote{
    In an alternative approach, instead of attributing the deviations
    from the free limit at smaller temperatures to the mass of the
    quasiparticles, a variable number of degrees of freedom is proposed
    in \cite{SchneW}.}
The existence of such heavy excitations, which we have inferred from
the thermodynamic bulk properties, has meanwhile been confirmed
directly by lattice calculations of the propagators \cite{Petre}.
Finally, since the derivative of the `bag' function $B$ is related to
the quasiparticle masses by Eq.~(\ref{B'}), the model is completely
defined by fixing
\be
  B_0 = B(T_c) = 1.1\, T_c^4 \, ,
\ee which enters the fit in Fig.~\ref{fig:latt dat p} as the third
parameter.

Since all the information about the coupling is encoded in the
parameters $T_s$ and $\lambda$, it is interesting to look at their
flavor dependence.  Comparing to the pure gauge plasma, it is
recalled that in this case the pressure becomes very small close to
the transition since there it has to match the pressure of the heavy
glue balls in the confined phase. Similarly, the entropy is small at
$T \sim T_c$, which implies a large coupling there. For $n_{\!f}=2$
the scaled entropy for $T \sim T_c$ is somewhat larger, thus close to
the transition the coupling has to be smaller than for pure SU(3).
However, for fixed parameters $\lambda$ and $T_s$, the coupling
(\ref{eq:alpha(T)}) would increase with increasing number of active
flavors.  Therefore, a difference of the parameters for $n_{\!f} = 2$
to those for the pure gauge plasma \cite{PKS00},
\be
 \lambda^{_{\rm SU(3)}} = 4.9 \, , \qquad
 T_s^{_{\rm SU(3)}} = 0.73\, T_c \, ,
\ee
is not unexpected.
Interestingly, the parameter $T_s$ does not change by much compared to
the case of $n_{\!f} = 2$.

\section{Nonzero chemical potential}

The quasiparticle model as applied in the previous section can be
generalized to nonzero quark chemical potential $\mu_q = \mu$. The
quasiparticle masses now depend also on $\mu$ -- explicitly by the
dimensionful coefficients of the coupling in Eq.~(\ref{m2}), and
implicitly by the coupling itself. As shown in \cite{PKS00},
Maxwell's relation, $\partial s / \partial \mu =
\partial n / \partial T$, directly implies a partial differential
equation for $\alpha_s(T,\mu)$. It is of first order and linear in
the derivatives of the coupling (but nonlinear in $\alpha_s$),
\be
   c_T \frac{\partial\alpha_s}{\partial T}
  +c_\mu \frac{\partial\alpha_s}{\partial\mu}
  =
  C \, ,
  \label{eq: flow eq}
\ee where the coefficients $c_T$, $c_\mu$ and $C$ depend on $T$,
$\mu$ and $\alpha_s$. It can easily be solved by reduction to a
system of coupled ordinary differential equations,
\be
  \frac{dT(s)}{ds} = c_T \, , \quad
  \frac{d\mu(s)}{ds} = c_\mu \, , \quad
  \frac{d\alpha_s(s)}{ds} = C \, ,
\ee which determines the so-called characteristic curves $T(s)$,
$\mu(s)$, and the evolution of $\alpha_s$ along such a curve, given
an initial value.

With regard to the underlying physics it is worth pointing out some
properties of the flow equation (\ref{eq: flow eq}). The coefficients
are combinations of products of a derivative of the quasiparticle
entropy or density with respect to the quasiparticle mass, and a
derivative of the quasiparticle mass with respect to $T$, $\mu$ or
$\alpha_s$. Writing down the explicit expressions, it is easy to see
that the flow equation is elliptic. In particular, one finds
\be
  c_T(T, \mu=0) = 0 \, , \qquad c_\mu(T=0, \mu) = 0 \, .
\ee The coefficient $c_\mu$, e.\,g., vanishes because not only the
entropy goes to zero as $T \rightarrow 0$, but also its derivative
with respect to the mass. Therefore, the characteristics are
perpendicular to both the $T$ and the $\mu$ axes. This guarantees
that specifying the coupling on some interval on the $T$ axis sets up
a valid initial condition problem. From the temperature dependence of
the effective coupling as obtained from the lattice data at $\mu =
0$, e.\,g.\ in the physically motivated parameterization
(\ref{eq:alpha(T)}), we can therefore determine numerically the
coupling from Eq.~(\ref{eq: flow eq}), and hence the equation of
state, in other parts of the $\mu\,T$ plane.

\begin{figure}[hbt]
  \centerline{\includegraphics[scale=1]{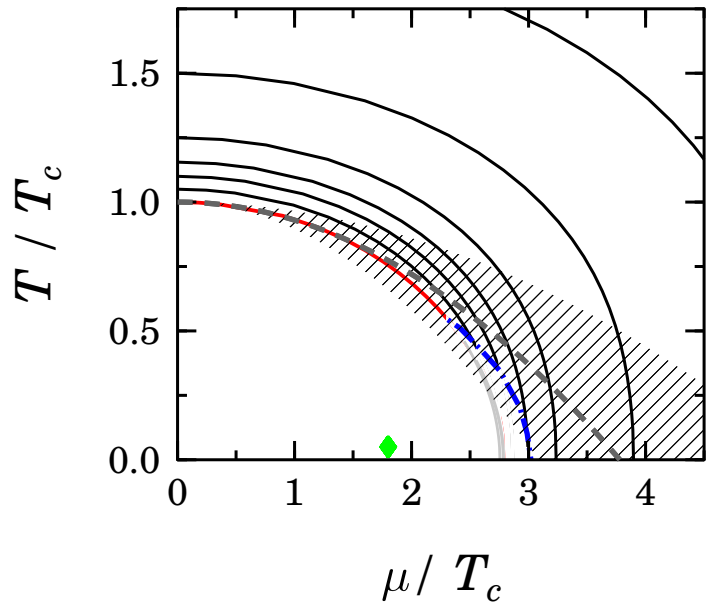}}
  \caption{Represented by the full lines are the characteristics of the
    flow equation (\ref{eq: flow eq}).
    The characteristic through $T_c$ coincides for small $\mu$ with the
    critical line (dashed, with a hatched error band) obtained in the
    lattice calculation \cite{AlltHKKLSS}.
    In the region under the dash-dotted line the resulting quasiparticle
    pressure is negative -- a transition to another phase has to happen
    somewhere outside.
    Therefore, the narrow grey region under the $p=0$ line, where the
    solution of the flow equation is not unique, is physically irrelevant.
    Indicated by the symbol (assuming, for the scaling, $T_c = 170\,$MeV)
    is the chemical potential $\mu_n$ in nuclear matter.
    \label{fig: char}}
\end{figure}

\begin{figure}[hbt]
  \centerline{\includegraphics[scale=1,angle=-90]{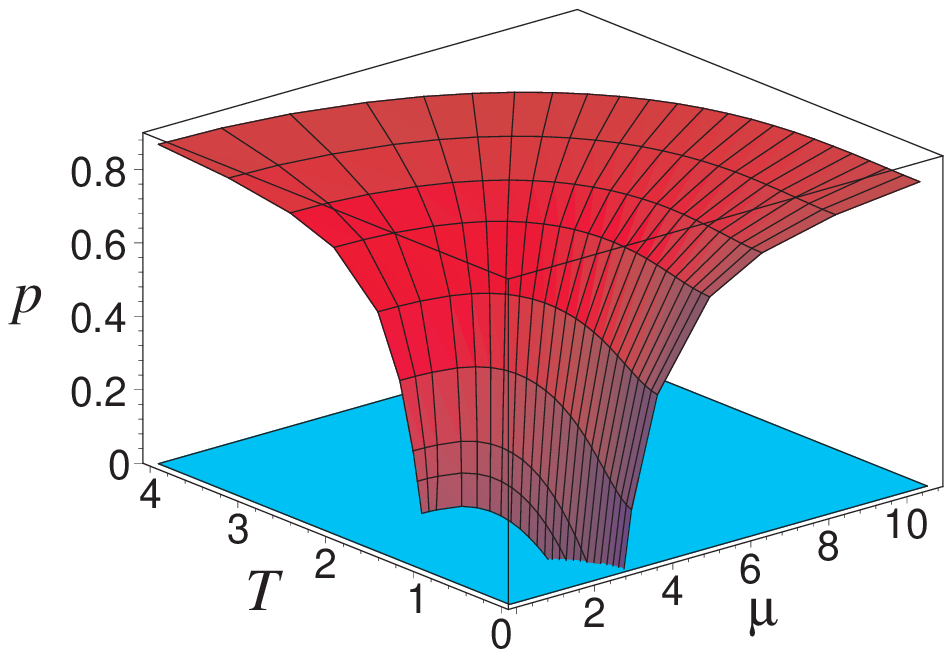}}
  \vspace*{-15mm}
  \caption{The pressure scaled by the free pressure $p_0(T,\mu)$;
    $T$ and $\mu$ are in units of $T_c$.
    The pressure along the characteristics starting out from $T \sim
    T_c$ becomes negative at small $T$, see also Fig.~\ref{fig: char}.
    The change to a different phase has to happen already outside this
    region.
    \label{fig: p_muT}}
\end{figure}

It is instructive to consider the asymptotic limit, $\alpha_s
\rightarrow 0$, of Eq.~(\ref{eq: flow eq}), where the coefficient
$C$ vanishes. Then the coupling is constant along the
characteristics, which become ellipses in the variables $T^2$ and
$\mu^2$, leading to the mapping
\be
  T
  \rightarrow
  \left( \frac{9n_{\!f}}{4N_c+5n_{\!f}} \right)^{\!1/4}\, \frac\mu\pi
  \; .
\ee This holds approximately also for larger coupling, see
Fig.~\ref{fig: char}, so the lattice data at $\mu=0$ are mapped in
elliptic strips into the $\mu\,T$ plane. On the other hand, an ansatz
analog to Eq.~(\ref{eq:alpha(T)}) to parameterize $\alpha_s(T=0,\mu)$ is
quantitatively less satisfactory than in the case $\mu=0$.
A closer look at the characteristics emanating from the interval
$[T_c,1.06T_c]$ reveals that they intersect in a narrow half-crescent
region, which indicates that there the solution of the flow equation
is not unique. This, however, is only an ostensible ambiguity. It so
happens that the extrapolation of the pressure becomes negative in a
larger region, see Figs.~\ref{fig: char} and \ref{fig: p_muT}. This
implies that a transition to another phase, at a certain positive
pressure, happens already outside this region, so the encountered
ambiguity of the flow equation is of no physical relevance.\footnote{
    We remark that the region where the solution of the flow equation
    is not unique is determined only by $\alpha_s(\mu=0,T)$, i.\,e.\ by
    the parameters $\lambda$ and $T_s$ fitted from the entropy, whereas
    the $p=0$ line depends also on $p(\mu=0,T_c)$ and thus on the
    third parameter $B_0$.
    Therefore, the fact that the potential ambiguity is irrelevant is
    based in a nontrivial way on the underlying lattice data for the
    equation of state at $\mu=0$.}

At this point we emphasize again that this extrapolation of the
quasiparticle model relies only on the requirement of thermodynamic
consistency. Of course, it implicitly assumes also that the
quasiparticle structure does not change, i.\,e., that deconfined
quarks and gluons are the relevant degrees of freedom. For small
enough $\mu$ and temperatures above (or near, as $\mu$ gets larger)
$T_c$ this is a justified assumption. However, the quasiparticle
structure will change in the hadronic phase, when both $T$ and $\mu$
are small, as well as for sufficiently cold and dense systems where
the color-superconducting phase is expected. Although the present
quasiparticle model cannot make any statements about these phases, it
is interesting to observe that it `anticipates' the existence of
another phase only from the lattice input at $T>T_c$ and $\mu=0$. An
interpretation of the apparent similarity of the line of vanishing
pressure in Fig.~\ref{fig: char} with the expected transition line
from the hadron to the superconducting quark matter phase, see
Ref.~\cite{RajaW}, remains, of course, a speculation.

There is, however, a related question which we can address with the
quasiparticle model without knowing details about the other phases,
just based on the fact that for nonzero chemical potential the
transition from the deconfined to the confined phase occurs at the
critical line $T_c(\mu)$.
The critical line is expected to be perpendicular to the $T$ axis, which
has been confirmed in a recent lattice calculation \cite{AlltHKKLSS} where
also its curvature at $\mu=0$ has been calculated\footnote{
    In passing we note the amusing fact that the result agrees with the
    value from the bag model assuming free massless pions for the hadronic
    phase.},
$T_c\, d^2 T_c(\mu)/d\mu^2|_{\mu=0} \approx -0.14$.
Within the quasiparticle model it is natural to relate, at least for small
$\mu$, the critical line to the characteristic through $T_c(\mu=0)$, which,
as shown above, is also perpendicular to the $T$ axis.
For small $\mu$ where only the quadratic terms are relevant (practically
even for $\mu$ as large as $2\,T_c$), we indeed find the $T_c$
characteristic in a striking agreement with the critical line from
\cite{AlltHKKLSS}, see Fig.~\ref{fig: char}.
Another argument supporting the above interpretation of the $T_c$
characteristic comes from considering the case where the quark flavors
have opposite chemical potentials, $\mu_u = -\mu_d = \tilde\mu$.
With this isovector chemical potential the fermion determinant is positive
definite, and standard Monte Carlo techniques can be applied to study this
system on the lattice \cite{AlfKW}.
The lattice result \cite{AlltHKKLSS} obtained for the curvature of the
critical line in that case agrees with the value quoted above for the
isoscalar potential $\mu$.
Within the quasiparticle model, the equality of these two numbers is
immediately evident.

In Ref.~\cite{AlltHKKLSS} it was furthermore mentioned that the
quadratic behavior, with the same curvature as at $\mu = 0$, of the
critical line is not likely to extrapolate down to small transition
temperatures since $T_c(\mu)$ would then vanish only at $\mu_c \sim
650\,$MeV. Phenomenologically, however, $\mu_c$ is expected to be not
very much larger, say at most by 200\,MeV, than the quark chemical
potential $\mu_n = 307\,$MeV in nuclear matter. In the quasiparticle
model, from the chemical potential where the extrapolated pressure
vanishes at $T=0$, we estimate $\mu_c \approx 3\,T_c \sim 500\,$MeV.

\begin{figure}[hbt]
  \centerline{\includegraphics[scale=1]{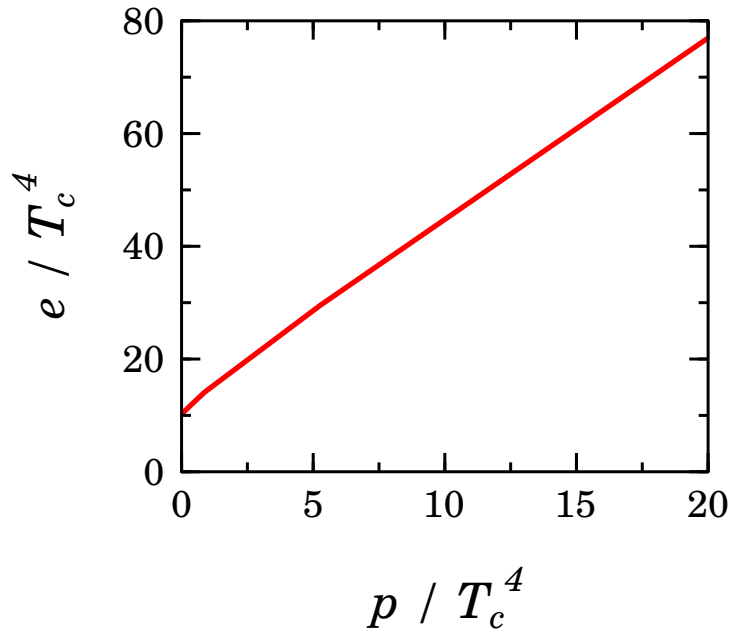}}
  \caption{The estimate for the equation of state of quark matter at
     $T=0$. \label{fig: e(p)}}
\end{figure}

This value is in the expected ball park, which encourages us to
consider the extrapolation of the model down to smaller temperatures.
Although for $T \rightarrow 0$ quark matter will be in the
superconducting phase, it is still possible to give an estimate of
its equation of state in that region from the quasiparticle model.
The quark pairing influences thermodynamic bulk properties at the
order of $(\Delta\,\mu)^2$, with the gap energy $\Delta$ being at
most 100\,MeV \cite{RajaW}. This has little effect on the energy
density $e = \sum_i e_i + B$ as both the quasiparticle contributions
and the function $B$ are parametrically of the order ${\cal
O}(\mu^4)$. For the pressure, on the other hand, the pairing effects
become comparable to our expression $p = \sum_i p_i - B$ only when
the latter becomes small. Since the pressure of the thermodynamically
favored superconducting phase is larger than that of the plasma
phase, the relation $e(p)$ as shown in Fig.~\ref{fig: e(p)} is
therefore an upper estimate of the equation of state of cold quark
matter.
For $p \ge 5\, T_c^4$, we obtain $e(p) \approx 13\, T_c^4 + 3.2\, p$,
where the slope is mainly determined by the fact that the pressure at
$T=0$ essentially scales as $\mu^4$. For smaller pressure, the slope
is only slightly larger, and the energy at $p=0$ is
approximately\footnote{
    This value renders more precisely the rough estimate \cite{PKS01},
    which was about 40\% larger. Based on the pragmatic extension
    of the continuum extrapolation of the lattice data \cite{Karsch00}
    shown in Fig.~\ref{fig:latt dat p} near $T_c$, the fit led to a
    similar value for $T_s$, but to $\lambda \approx 11$.
    This demonstrates that details of the underlying lattice data are
    important for quantitative predictions at $\mu \not= 0$ but, on
    the other hand, that the estimates are rather robust.}
$11\, T_c^4$. Assuming $T_c \approx 170\,$MeV, this translates into
an energy density of 1\,GeV/fm$^3$. Bearing in mind that this is an
upper estimate, and comparing to the bag model equation of state,
$e^{\rm bag}(p) = 4\tilde B + 3p$, this result is still considerably
larger than estimates with commonly assumed values of the bag
constant $\tilde B$.

Coming back to the region of the phase space where the quasiparticle model
is well grounded, we finally address the question of the behavior of the
pressure and the energy density along the critical line near $\mu=0$.
In the lattice simulations \cite{AlltHKKLSS} both quantities have been
found to be constant within the numerical errors.
This is compatible with our result for small $\mu$,
\be
  p(T_c(\mu),\mu)-p(T_c(0),0) \approx -0.02\, \mu^2\, T_c^2 \, .
\ee
The corresponding change in the energy density is about three times
larger.
These results differ notably from the estimate from the bag model
which, although the critical line has a similar shape for small $\mu$,
would yield coefficients larger by a factor of four.

\section{Conclusions}

Within our quasiparticle model \cite{PKPS96,PKS00} we analyze recent
$n_{\!f} = 2$ QCD lattice calculations \cite{CPPACS} of the equation
of state at nonzero temperature and $\mu = 0$, and then extend the
quasiparticle model to nonzero baryon density. The resulting elliptic
flow equation for the coupling relates the thermodynamic potential
along the characteristic curves in the $\mu\,T$ plane. We argue that
the characteristic line through $T_c(\mu=0)$ is related to the
critical line in the phase diagram. This is confirmed by comparing
our results for the curvature of the critical line at $\mu = 0$, and
the variation of the equation of state along it, with recent lattice
simulations \cite{AlltHKKLSS} exploring the region of small $\mu$.

We give an estimate for the equation of state of cold quark matter.
Energy density and pressure are almost linearly related, as in the
bag model, however with parameters obtained from the lattice data at
$\mu=0$. The relevant physical scale is given by the transition
temperature $T_c$, and the parameter corresponding to the bag
constant turns out to be large compared to conventional estimates,
$\lsim 250\,$MeV$^4$.

We have restricted ourselves to the case $n_{\!f}=2$, for which the
lattice data for the equation of state at $\mu = 0$ appear to be
established best. However, we expect similar results for other
numbers of flavors since the pronounced decrease of the ratio $p/p_0$
as $T$ approaches $T_c$, which indicates a large coupling strength,
seems to be generic. This universality is then echoed at nonzero
$\mu$ because for all $n_{\!f}$ the flow equation behaves for strong
coupling similarly as in the perturbative limit, where $\alpha_s$ is
constant along the elliptic-like characteristics. Indeed, the shape
of the phase boundary calculated in \cite{FodorK} for the physically
relevant case $n_{\!f}=2+1$, although now being a crossover near
$T_c$, is very similar to the shape for $n_{\!f}=2$. With the same
reasoning, we remark that our estimates are robust with respect to
remaining uncertainties of the underlying lattice data. Indeed, the
equation of state at $\mu \not= 0$ is not very sensitive to the
precise values of the model parameters as long as they reasonably
describe the gross features of the equation of state at $\mu = 0$.
Therefore, the large energy density at small pressure seems to be a
general feature of the equation of state.

As shown in \cite{PKS00,PKS01}, this would allow for pure quark stars
with masses $\le 1 M_\odot$ and radii $\le 10$ km. Similar small and
light quark stars have also been obtained within other approaches,
cf.~\cite{Blasch}. Such objects are of interest in the ongoing
discussion of the data of the quark star candidate RXJ1856.5-3754
\cite{Pons}. It should be emphasized, however, that the outermost
layers of such pure quark stars are metastable with respect to
hadronic matter with a larger pressure at $\mu \sim \mu_c$. The
details of the star structure depend sensitively on the hadronic
equation of state \cite{BK}. However, as discussed in \cite{Fraga}, a
stable branch of hybrid stars with a dense quark core and a thin
hadronic mantle could indeed be possible.
\newpage\noindent
{\bf Acknowledgments:} We would like to thank M.~Alford, E.~Fraga,
R.~Pisarski, and D.~Rischke for discussions.
This work is supported by BMBF.

\end{document}